\documentstyle[editedvolume]{crckapb}

\def\aj#1{{\it Astron. J.,} {\bf #1}}
\def\apj#1{{\it Astrophys. J.,} {\bf #1}}

\def\aa#1{{\it Astron. \& Astrophys.,} {\bf #1}}

\def\mn#1{{\it Mon. Not. Roy. Astr. Soc.,} {\bf #1}}

\def\ass#1{{\it Astrophys. \& Sp. Sci.,} {\bf #1}}

\def\be{\begin{equation}}
\def\ee{\end{equation}}

\def\bea{\begin{eqnarray}}
\def\eea{\end{eqnarray}}

\def\ie{{\it i.e.}}
\def\etal{{\it et al.}}

\begin{opening}

\title{Chemo -- Dynamical evolution of disk galaxies, \protect \\
       smoothed particles hydrodynamics approach}
       
\subtitle{Chemo -- Dynamical SPH code}

\author{Peter Berczik}
         
\institute{Main Astronomical Observatory of \\
           Ukrainian National Academy of Sciences \\
           252650, Golosiiv, Kiev-022, Ukraine \\
           e-mail: {\tt berczik@mao.kiev.ua}}
           
\end{opening}

\runningtitle{Chemo -- dynamical SPH code}

\begin{document}


\begin{abstract}

     A new Chemo -- Dynamical Smoothed Particle Hydrodynamic (CD -- SPH) 
code is presented. The disk galaxy is described as a multi -- fragmented 
gas and star system, embedded into the cold dark matter halo. The star 
formation (SF) process, SNII, SNIa and PN events as well as chemical 
enrichment of gas have been considered within the framework of standard 
SPH model. Using this model we try to describe the dynamical and 
chemical evolution of triaxial disk -- like galaxies. It is found that 
such approach provides a realistic description of the process of 
formation, chemical and dynamical evolution of disk galaxies over the 
cosmological timescale.

\end{abstract}


\section{The CD -- SPH code and results}

     The simplicity and numerical efficiency of the SPH method were the 
main reasons why we chose similar technique for the modelling of the 
evolution of complex, multi -- fragmented triaxial protogalactic 
systems. We used our own modification of the hybrid N -- body/SPH method 
\cite{BerK96a,BerK96b,BerK97,Ber98}, which we call as chemo -- 
dynamical SPH (CD -- SPH). The "dark matter" and "stars" were included 
into the standard SPH algorithm as the N -- body collisionless system of 
particles, which can interact with the gas component only through the 
gravitation \cite{K92}. The star formation process and supernova 
explosions were included into the scheme in the manner proposed by 
\cite{RVN96,CLC97} but with our own modifications. 

\begin{figure*}[htbp]

\vspace{4.0cm}
\includegraphics{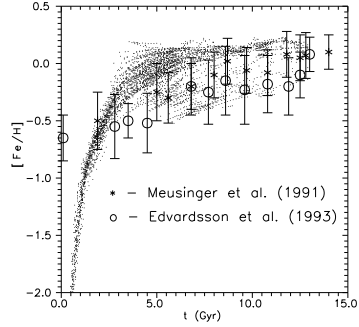}

\caption{[Fe/H]$(t)$. The age metalicity relation of the "star" particles in the 
"solar" cylinder ($ 8 $ kpc $ < \; r \; < \; 10 $ kpc)}
 
\label{fig-chem_1}
\end{figure*}

     For description of the process of the gas material convertion into 
stars we modify the standard SPH star formation algorithm 
\cite{K92,NW93} taking into account the presence of chaotic motions in 
the gaseous environment and the time lag between initial development of 
suitable conditions for star formation and star formation itself 
\cite{Ber98}. After the formation, these "star" particles return the 
chemically enriched gas to surrounding "gas" particles due to SNII, SNIa 
and PN events. For the description of this process we use the 
approximation proposed by \cite{RVN96}. We consider only the production 
of $^{16}$O and $^{56}$Fe, and try to describe the full galactic time 
evolution of these elements, from the beginning to present time (\ie~ $ 
t_{evol} \approx 13.0 $ Gyr). 

     In our calculations, as a first approximation, it is assumed that 
the model galaxy halo contains the CDMH component with Plummer -- type 
density profiles \cite{DC95}.

\begin{figure*}[htbp]

\vspace{4.0cm}
\includegraphics{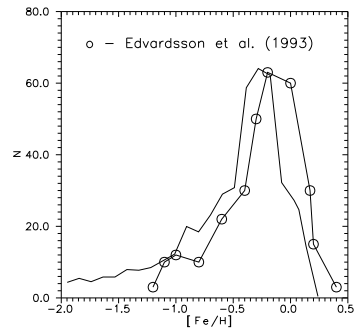}

\caption{$N_{*}($[Fe/H]$)$. The metallicity distribution of the "star" particles 
in the "solar" cylinder ($ 8 $ kpc $ < \; r \; < \; 10 $ kpc)}
 
\label{fig-chem_2}
\end{figure*}

     The SPH calculations were carried out for  $ N_{gas} = 2109 $ "gas" 
particles. According to \cite{NW93,RVN96}, such number seems to be quite 
enough to provide qualitatively correct description of the system 
behaviour. Even such small number of "gas" particles produces a $ 
N_{star} = 31631 $ "star" particles at the end of calculation. 

     As initial model (relevant for CDM -- scenario) we took constant -- 
density homogeneous gaseous triaxial configuration ($ M_{gas} = 10^{11} 
\; M_\odot $) within the dark matter halo ($ M_{halo} = 10^{12} \; 
M_\odot $). We set $ A = 100 $ kpc, $ B = 75 $ kpc and $ C = 50 $ kpc 
for semiaxes of system. Such triaxial configurations are reported in 
cosmological simulations of the dark matter haloes formation 
\cite{EL95,FWDE88,WQSZ92}. We set the smoothing parameter of CDMH: $ 
b_{halo} = 25 $ kpc. 

\begin{figure*}[htbp]

\vspace{4.0cm}
\includegraphics{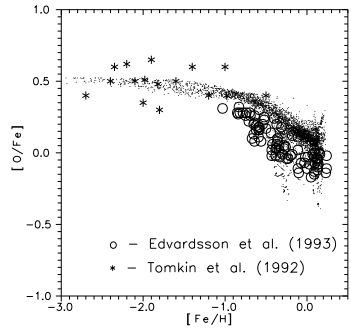}

\caption{The [O/Fe] vs. [Fe/H] distribution of the "star" particles in the 
"solar" cylinder ($ 8 $ kpc $ < \; r \; < \; 10 $ kpc)}
 
\label{fig-chem_3}
\end{figure*}

\begin{figure*}[htbp]

\vspace{4.0cm}
\includegraphics{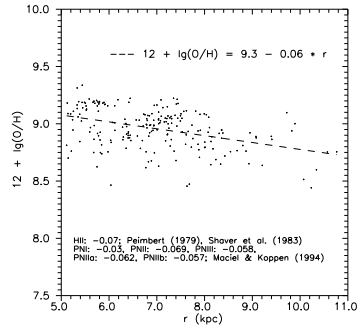}

\caption{[O/H]$(r)$. The [O/H] radial distribution}
 
\label{fig-grad-z}
\end{figure*}

     To check the SF and chemical enrichment algorithm in our SPH code, 
we use the chemical characteristics of the disk in "solar" cylinder ($ 8 
$ kpc $ < \; r \; < \; 10 $ kpc). The age metalicity relation of the 
"star" particles in the "solar" cylinder [Fe/H]$(t)$ we present in 
Fig.~\ref{fig-chem_1}. We presented in Fig.~\ref{fig-chem_2}. the 
metallicity distribution $N_{*}($[Fe/H]$)$. The [O/Fe] vs. [Fe/H] 
distribution we presented in Fig.~\ref{fig-chem_3}. In figures we also 
present the observational data from \cite{MRS91,EAG93,TLLS92}. The [O/H] 
radial distribution [O/H]$(r)$ we presented in Fig.~\ref{fig-grad-z}. At 
the distances $ 5 $ kpc $ < \; r \; < \; 11 $ kpc the model radial 
abundance gradient is $ -0.06 $ dex/kpc. Our model has a very good 
agreemnet with oxygen radial gradient in the our Galaxy 
\cite{P79,SMNDP83}.

\subsection{Conclusion}

     This simple model provides a good, self -- consistent picture of 
the process of galaxy formation and star formation in the galaxy. The 
dynamical and chemical evolution of modelled disk -- like galaxy is 
coincident with the results of observations for our own Galaxy. The 
results of our modelling give a good base for a wide use of proposed SF 
and chemical enrichment algorithm in other SPH simulations.


\bigskip

     {\bf Acknowledgements:} The author is grateful to S.G. Kravchuk, 
L.S. Pilyugin and Yu.I. Izotov for fruitful discussions during the 
preparation of this work. This research was supported by a grant from 
the American Astronomical Society.



\end{document}